\documentstyle[aps]{revtex}

\newcommand{\greeksym}[1]{{\usefont{U}{psy}{m}{n}#1}}
\newcommand{\umu}{\mbox{\greeksym{m}}}

\newcommand{\0}{|0\rangle}
\newcommand{\1}{|1\rangle}
\newcommand{\ad}{a^{\dagger}}
\newcommand{\ve}{\mbox{\greeksym{e}}}
\newcommand{\Yb}{$^{171}$Yb$^+$}

\begin{document}
\title{Conditional Spin Resonance with Trapped Ions\protect\footnote{in {\it 
Laser Physics at the Limit}, Springer, Heidelberg, 2001, p.261. 
Dedicated to T. W. H\"ansch on the occasion of his 60th 
birthday.}}

\author{Christof Wunderlich}

\address{Institut f\"ur Laser-Physik, Universit\"at Hamburg, Jungiusstr. 9, 
20355 Hamburg, Germany}
\date{6 August 2001}
\maketitle
\begin{abstract}
Internal states of different ions in an electrodynamic trap are 
coupled when a static magnetic field is applied -- analogous to 
spin-spin coupling in molecules used for NMR. This spin-spin 
interaction can be used, for example, to implement quantum logic 
operations in ion traps using NMR methods. The collection of 
trapped ions can be viewed as a $N$-qubit molecule with 
adjustable coupling constants.
\end{abstract}

\section{Motivation}

Digital information processing builds upon elementary physical
elements (``bits") that may occupy either one of two possible
states labelled 0 and 1, respectively. If a quantum system, for
example, an individual atom having discrete energy eigenstates, is
chosen as elementary switch (``qubit" ), then the general state 
of this system will be a superposition of the two  computational 
basis states, i.e. the states chosen to represent the logic 0 and 
1. When applying the superposition principle to a register 
comprising $N$ qubits, one immediately sees that such a register 
can exist in a superposition of $2^N$ states thus representing 
$2^N$ binary encoded numbers simultaneously. Any operation on 
this register will act on all states at once, effecting parallel 
processing on an exponentially growing (with $N$) number of 
states. The outcome of a measurement on this register after such 
an operation will, of course, yield just one out of $2^N$ 
possible results with a certain probability. 

In order to take advantage of quantum
parallelism for efficient computing, a second ingredient is
necessary: interference. A useful quantum algorithm has to exploit
this parallelism, and, at the same time, make different
computational paths interfere such that only the correct result
survives after the last computational step \cite{Cleve98}. An
important example is Shor's algorithm for the factorization of
large numbers \cite{Shor94}. Once created, coherent superpositions
have to remain intact while a quantum algorithm is carried out,
i.e. qubits must not in an uncontrollable way interact with their
environment. This would lead to decoherence, an important issue,
not only in the realm of quantum information processing (QIP), but
also related to the notion of measurement in quantum mechanics
\cite{Brune96,Toschek01}.

A quantum computer is ideally suited for the simulation of quantum
mechanical systems \cite{Feynman82,Molmer00}, for example, to
determine eigenvalues and eigenvectors of many-body systems
\cite{Abrams99}. Calculating the dynamics of chaotic systems is
another useful line of action for a quantum computer, even for one
that consists of only a few qubits \cite{Georgeot01A}. Beneficial
both for fundamental research and applications is the ability of a
quantum computer -- comprising a modest number of qubits and
working with limited accuracy -- to simulate the dynamics of a
macroscopic ensemble of classical particles, a task not suitable
even for modern supercomputers \cite{Georgeot01}.

In the course of a quantum computation entangled states of qubits
are created exhibiting correlations between individual qubits that
possess no classical analog. Fundamental questions concerning the 
role of entanglement, not only in QIP, but also in the framework 
of general physics \cite{Brukner01} add more motivation to 
exploring the field of QIP. In 1935 Einstein, Podolsky, and Rosen 
scrutinized quantum mechanical predictions for two entangled 
particles and found non-local correlations between these 
particles \cite{Einstein35}. This, what Einstein called, \lq 
spooky action at a distance' prompted him to call into question 
quantum theory. During the last decade various experiments 
succeeded in preparing and analyzing entangled states of 
different physical systems \cite{Hagley97,Turchette98}, which 
marked the beginning of controlled manipulation of entanglement 
of massive particles. On the theoretical side, too, the search 
for better understanding, quantification, and use of entanglement 
as a resource for QIP is a very active field \cite{Lewenstein00}.

\section{Trapped ions and QIP}

QIP is an interdisciplinary field of research, whose results will
have significant impact both on basic research and applied
sciences. Theory in this field is still well ahead of experimental
progress and manageable experimental systems are needed. Essential
characteristics of a device designed for quantum computing include
\cite{DiVincenzo00} the scalability of the system, the ability to
reset the qubits' states to a known one, and to make qubit
specific measurements. Furthermore, decoherence times have to be
much longer than the typical gate operation time. Finally, a set
of quantum gates is needed to construct any desired unitary
transformation of $N$ qubits. A sequence of unitary
transformations that make up a quantum algorithm can be broken
down into two operational elements sufficient for the synthesis of
any quantum algorithm \cite{DiVincenzo95}: i) the preparation of
individual qubits in arbitrary superposition states, and ii) the
execution of conditional dynamics on different qubits, which is at
the heart of quantum computing. It is this last requirement we
will be mainly concerned with in this chapter.

A promising system for QIP are electrodynamically trapped ions
where two internal states of each ion, labelled $\0$ and $\1$ in
the remainder of this chapter, are chosen as one qubit
\cite{Cirac95}. Conditional dynamics with $N$ trapped ions require
coupling of their internal and external degrees of freedom.
Following the first preparation and detection of a single atom
reported in \cite{Neuhauser80} -- prerequisite for many important
studies with trapped ions -- principal elements of ion trap
quantum computing have been realized experimentally (for instance,
\cite{King98,Peik99,Roos99,Huesmann99,Sackett00,Reiss01}.)

Ted H\"{a}nsch once illustrated  the principles of electrodynamic
trapping using macroscopic charged particles \cite{Wuerker59}.
After testing various kinds of electrode configurations he finally
arrived at the ultimate simplification: a conventional paper clip,
connected to a regular power socket sufficed to stably trap
charged lycopodium seeds.  He documented his efforts with a
humorous video that, for example, shows the periodic motion of
particles in step with ballet music. This may shed a little light
on Ted H\"{a}nsch's imaginative, playful approach to physics that
enabled him to make so many outstanding contributions.

The vibrational motion of a collection of ions (the \lq \lq bus
-qubit'') is used as means of communication between individual
qubits to implement conditional quantum dynamics in ion traps
\cite{Cirac95}. Thus, initial cooling of the ions' motional  
degrees of freedom is indispensable for QIP. Optical cooling of 
atoms, suggested by Ted H\"{a}nsch and Arthur Schawlow 
\cite{Haensch75} and for trapped atoms by Wineland and Dehmelt 
\cite{Wineland75}, has for the first time been observed on a 
collection of trapped ions \cite{Neuhauser78}.

\subsection{Why is optical radiation used ?}
Common to all experiments -- related either to QIP or other
research fields -- that require some kind of coupling between
internal and external degrees of freedom of atoms is the use of
optical radiation for this purpose. The parameter determining the
coupling strength between internal and motional dynamics is the
so-called Lamb-Dicke parameter
\begin{equation}
\eta \equiv \sqrt{\frac{(\hbar k)^2}{2m} / \hbar \nu_1} = \Delta
             z_1\, k \label{LDP}
\end{equation}
the square of which gives the ratio between the change in kinetic
energy of the atom due to the absorption or emission of a photon
and the quantized energy spacing of the harmonic trapping
potential characterized by the angular frequency $\nu_1$ ($k$ is
the wave vector of the light field, $m$ the mass of the atom, and
$\Delta z_1= \sqrt{\hbar/2m\nu_1}$ signifies the spatial extent of
the vibrational ground state wave function of the atom). Only if
$\eta$ is nonvanishing will the absorption or emission of photons 
be possibly accompanied by a change of the motional state of the 
atom. This is apparent when the Hamiltonian describing the 
coupling between an applied electromagnetic field of angular 
frequency $\omega$ and a harmonically trapped 2-state atom is 
considered: 
\begin{equation}
H_I = \frac{1}{2} \hbar\Omega_R (\sigma_+ + \sigma_-)
\left[\exp[i(\eta (\ad + a)-\omega t + \phi')] +
\mbox{h.c.}\right] \; ,
\end{equation}
where $\Omega_R= \vec{d}\cdot\vec{F} / \hbar$ is the Rabi
frequency with $\vec{d}\cdot\vec{F}$ signifying either magnetic or
electric coupling between the atomic dipole and the respective
field component. $\sigma_{+,-} = 1/2 \;(\sigma_x \pm \sigma_y )$
are the atomic raising and lowering operators, respectively,
$\Delta z_1 (\ad +a)$ is the position operator, and $\phi'$ is the
initial phase of the driving field. Trapping a $^{171}$Yb$^+$ 
ion, for example, with $\nu_1 = 2\pi \, 100$kHz gives $\Delta z_1 
\approx 17$nm and \ref{LDP} shows that driving radiation in the 
optical regime is necessary to couple internal and external 
dynamics of trapped atoms.

Here, and in the remainder of this article, we consider a Paul
trap \cite{Paul58} in a linear configuration where a 
time-dependent two-dimensional quadrupole field strongly confines 
the ions in the radial direction yielding an average effective 
harmonic potential \cite{Ghosh}. An additional static electric 
field is chosen such that the ions are harmonically confined also 
in the axial direction \cite{Prestage89}. If the confinement of 
$N$ ions is much stronger in the radial than in the axial 
direction, the ions will form a linear chain \cite{Schiffer93} 
with typical inter-ion distance $\delta\!z = \zeta\, 2N^{-0.57}$ 
where
 $\zeta\equiv (e^2/4\pi\epsilon_0 m \nu_1^2)^{1/3}$
\cite{Steane97}. The distance between neighboring ions $\delta\!z$
is determined by mutual Coulomb repulsion of the ions and trap
frequency $\nu_1$ in the axial direction. Manipulation of
individual ions is achieved by focusing laser light to a spot size
smaller than $\delta\!z$. Typically, $\delta\!z$ is of the order
of a few $\umu$m; for example, $\delta\!z\approx 7 \umu$m for
$N=10$ $^{171}$Yb$^+$ ions with $\nu_1=2\pi \, 100$kHz. Again,
only optical radiation is useful for this purpose.

\subsection{Spin resonance}
Many phenomena that were only recently studied in the optical
domain form the basis for techniques belonging to the standard
repertoire of coherent manipulation of nuclear and electronic
magnetic moments associated with their spins. One reason for the
tremendous and fast success of nuclear magnetic resonance (NMR)
experiments in the field of QIP is the high level of
sophistication that experimental techniques in this field have
reached over decades. This is an impressive example for a
successful technology whose basis was developed by physicists
\cite{Rabi39} and that has overcome the boundaries between
disciplines of science. For many years researchers, for example,
in chemistry and in the life sciences have routinely used
commercial NMR equipment. The technological basis for NMR -- apart
from the preparation of the samples to be investigated -- is the
generation and coherent manipulation of electromagnetic radiation
in the radiofrequency (rf) and microwave (mw) regime. This
treasure of knowledge and technology could immediately be
exploited, again for fundamental research, in the emerging field
of QIP, where even complete algorithms based on quantum logic have
been demonstrated \cite{Jones98,Chuang98}.

There are also drawbacks associated with NMR quantum computing,
for example, considerable effort has to be devoted to the
preparation of pseudo-pure states of a macroscopic ensemble of
spins with initial thermal population distribution. This
preparation leads to an exponentially growing cost (with the
number $N$ of qubits) either in signal strength or the number of
experiments involved \cite{Vandersypen00}, since the fraction of
spins in their ground state is proportional to $N/2^N$. Extending
NMR quantum computing to larger numbers of qubits than in present
experiments will also require molecules with more nuclear spins
distinct in their resonance frequencies and, at the same time,
with appreciable coupling constants.

Trapped ions, on the other hand,  provide individual qubits -- for
example hyperfine states -- well isolated from their environment.
However, the application of mw radiation for quantum logic
operations with a string of ions is not possible, since i) this
long wavelength radiation does not couple internal and external
degrees of freedom of the ions, and ii) focusing down to the
required small spot sizes for access to individual qubits is not
possible. It would be desirable to combine the advantages of
trapped ions and NMR techniques in future experiments.

\section{A modified ion trap}

An axial magnetic field gradient applied to an electrodynamic trap
indeed has the desired effect of coupling internal state dynamics
and motion of the ions when mw driving radiation is applied
\cite{Mintert01}. In addition, the field gradient serves to
separate qubit resonances of individual ions making them
distinguishable in frequency space. Thus microwave radiation can
be used to coherently manipulate hyperfine states of individual
ions and condition their internal dynamics on the states of other
qubits. The treatment put forward in \cite{Mintert01} is
generalized in what follows and it is shown that mutual spin--spin
coupling between qubits arises in such a modified ion trap
analogous to the coupling Hamiltonian in molecules used for NMR.
The size of this NMR-type coupling is proportional to the square
of the ratio between magnetic field gradient $\partial_z B$ and
$\nu_1$.

The non-relativistic Hamiltonian describing the internal dynamics
of a diatomic molecule may be written as \cite{Lefebvre}
\begin{equation}\label{Molekuel}
 H_{M}= T_{N} + T_{el} + V(\vec{r},R)
\end{equation}
where $T_{N}$ and $T_{el}$ represent the kinetic energy operator
of nuclear and electronic motion, respectively. All electrostatic
potential energy terms are contained in $V(\vec{r},R)$, with
$\vec{r}$ denoting the collection of electronic coordinates and
$R$ the internuclear distance. Neglecting initially the nuclear
kinetic energy yields the Schr\"odinger equation for the
electronic wave functions
\begin{equation}\label{BO}
  (T_{el} + V(\vec{r},R)) \Phi_a(\vec{r},R) \equiv
  H_{el}\Phi_a(\vec{r},R) = E_{el,a}(R)\Phi_a(\vec{r},R)\; .
\end{equation}
These Born-Oppenheimer (BO) wave functions depend on $R$ as a 
parameter. With $\langle\Phi_a|T_{N}|\Phi_a\rangle \chi \approx 
T_{N}\langle\Phi_a|\Phi_a\rangle \chi = T_{N} \chi$ the 
Schr\"odinger equation for the nuclear motional wave functions 
$\chi$
\begin{equation}
 (T_N + E_{el,a}) \chi = E_T \chi
\end{equation}
determines the dynamics of the nuclei on the BO potential energy
curves $E_{el,a}$.

We now turn to the description of a linear chain of $N$ 
harmonically trapped, singly ionized two-level ions in an 
analogous way. The electronic part of the total Hamiltonian can 
be solved independently for each ion, since the distance 
$\delta\!z$ between different ions is much larger than the extent 
of individual spatial wave functions. Two Zeeman states, 
$|E_{0n}\rangle$ and $|E_{1n}\rangle$, of each ion serve as one 
qubit ($n=1,\ldots ,N$). The overall electronic state of the ions 
obeys $H_{el}\Phi_a(\vec{z}) = E_{el,a}(\vec{z})\Phi_a(\vec{z})$ 
with
\begin{equation}\label{Hel}
  H_{el}=\frac{1}{2}\hbar \sum_{n=1}^N \omega_n(z_n)\sigma_{z,n}
\end{equation}
and  $\Phi_a(\vec{z})= \prod_{n=1}^N |E_{cn}(z_n)\rangle$, where 
$a=1\ldots 2^N$, $c=0,1$; $z_n$ denotes the axial coordinate of 
ion $n$, and $\sigma_z$ is the usual Pauli matrix. The qubit 
transition frequency $\omega_n = (E_{1n} - E_{0n})/\hbar$. A 
magnetic field applied to the linear arrangement of ions shifts 
the qubit states $|E_{cn}\rangle$ depending on the location $z_n$ 
of the $n-$th ion (here, $\vec{B}=bz\cdot\hat{z}+B_0$ is assumed 
for clarity, with $\hat{z}$ being the unit vector in the axial 
direction). The complete Hamiltonian for the ion chain is given by
\begin{eqnarray}\label{Hamilton}
H & = &  H_{el}(\vec{z}) + T_{A}(\vec{z}) + V_{A}(\vec{z})\nonumber \\
  & = &  H_{el}(\vec{z})
        + \frac{1}{2m}\sum_{n=1}^N p_{z,n}^2
        + \frac{m}{2}\sum_{n=1}^N \nu_1^2 z
        + \frac{e^2}{8\pi\epsilon_0}\sum_{n\neq l}^N \frac{1}{|z_n - z_l|}
\end{eqnarray}
The potential energy relevant for the motion of the ions is
obtained from
 $\langle\Phi_a|(H_{el} + V_A(\vec{z}))|\Phi_a\rangle
 = E_{el,a} + V_A(\vec{z})$. When there is no field gradient
present, i.e. $b=0$, the electronic energy is independent of $z$
and simply gives an additive constant. Therefore, only $T_A$ and
$V_A$ have to be considered in this case. Expanding $V_A$ around 
the equilibrium positions $z_{0,n}$ of the ions in terms of $q_n 
\equiv z_n - z_{0,n}$ up to second order yields the dynamical 
matrix $\hat{A}$ with $A_{ln}\equiv
\partial_{z_l} \partial_{z_n} V_A$ and the Hamiltonian of a
harmonic oscillator is obtained
\begin{equation}\label{HO}
T_A + V_A = \frac{1}{2m}\sum_{n=1}^N P_{Q,n}^2
        + \frac{m}{2}\sum_{n=1}^N \nu_n^2 Q_n
\end{equation}
with $N$ uncoupled vibrational modes \cite{James98}. The normal
coordinates $\vec{Q}$ and local coordinates $\vec{q}$ are
connected via $\vec{q}=\hat{S}\vec{Q}$ where $\hat{S}$ is the
unitary transformation matrix that diagonalizes $\hat{A}$.
Further, $P_{Q,n}=m\dot{Q}_n$.

Taking into consideration the field gradient, a new term in the
potential energy arises for ion $j$:
\begin{eqnarray}\label{V_B}
 \langle\Phi_a|H_{el,j}(\vec{z})|\Phi_a\rangle
 &=& E_{cj}(z_{0,j}) +
   \underbrace{\frac{\hbar}{2} \left.\frac{\partial\omega_j}{\partial z_j}\right|_{z_{0,j}}
   \,q_j (-1)^{c+1}}_{V_B}  \; .
\end{eqnarray}

An order of magnitude estimate of the size of the additional
potential energy term $V_B$ experienced by ion $j$, is obtained
upon substitution of $q_j \approx \Delta z_1$ into \ref{V_B}. The
new term, $V_B$ has to be compared to $\hbar \nu_1$, the ground 
state energy of the unperturbed lowest oscillator mode:
\begin{equation}\label{coupling}
  \ve \equiv \frac{|V_B|}{\hbar \nu_1} =
  \frac{|\partial_z\omega_j|\,\Delta z_1}{\nu_1}\; .
\end{equation}
As long as $\ve$ is much smaller than unity, the eigenfrequencies
of the oscillator modes only negligibly depend on the additional
potential term introduced by the Zeeman shift of the ionic qubit
states. Therefore, the part of the Hamiltonian that describes the
motional state of the ion string is well approximated by the
unperturbed harmonic oscillator, and the complete Hamiltonian
reads
\begin{eqnarray}
 H &=& \frac{\hbar}{2} \sum_{n=1}^N \omega_n(z_{0,n})\sigma_{z,n} +
       \frac{1}{2m}\sum_{n=1}^N P_{Q,n}^2 +
       \frac{m}{2}\sum_{n=1}^N \nu_n^2 Q_n^2  \nonumber \\
   & & \mbox{} + \frac{\hbar}{2}\sum_{n=1}^N\left[
       \left.\frac{\partial\omega_n}{\partial z_n}\right|_{z_{0,n}}
         \sigma_{z,n}\sum_{l=1}^N S_{ln}Q_l\right] \nonumber \\
   &=& \frac{\hbar}{2} \sum_{n=1}^N \omega_n(z_{0,n})\sigma_{z,n} +
       \frac{1}{2m}\sum_{n=1}^N P_{Q,n}^2 \nonumber \\
   & & \mbox{} + \frac{m}{2}\sum_{l=1}^N \nu_l^2
        \left[Q_l + \frac{\hbar}{2m\nu_l^2}\sum_n \left.\frac{\partial\omega_n}{\partial
        z_n}\right|_{z_{0,n}}\sigma_{z,n} S_{ln}\right]^2 \nonumber \\
   & & \mbox{} - \underbrace{\frac{\hbar}{4m}\sum_{l=1}^N \frac{1}{\nu_l^2}
        \left[\sum_n \left.\frac{\partial\omega_n}{\partial z_n}\right|_{z_{0,n}}
        \sigma_{z,n} S_{ln}\right]^2}_{H_{SS}}
\end{eqnarray}
with the electronic energy expanded up to first order in $q_n$.
The unitary transformation $\tilde{H}= U^\dagger HU$ with
\begin{equation}\label{U}
  U = \exp\left[
            -i\sum_l\left(
                    \frac{1}{2m\nu_l^2}
                    \sum_n\left.\frac{\partial\omega_n}{\partial z_n}\right|_{z_{0,n}}
                    \sigma_{z,n} S_{ln} \right)P_{Q,l}
           \right]
\end{equation}
yields
\begin{eqnarray}\label{HTilde}
  \tilde{H}&=& \frac{\hbar}{2} \sum_{n=1}^N \omega_n(z_{0,n})\sigma_{z,n} +
             \sum_{n=1}^N \frac{P_{Q,n}^2}{2m} +  \frac{m}{2}\nu_n^2
             Q_n^2 - H_{SS} \; .
\end{eqnarray}
Expressing the harmonic oscillator in \ref{HTilde} in terms of
creation and annihilation operators $a_n^\dagger$ and $a_n$,
respectively, using the definitions
\begin{eqnarray}
 \ve_{nl}&\equiv & S_{nl}\frac{\partial_z\omega_l\Delta z_n}{\nu_n}\; , \\
 J_{nl} &\equiv & \sum_{j=1}^N
                              \nu_j\ve_{jn}\ve_{jl} \label{J}\; ,
\end{eqnarray}
and after dropping constant terms, \ref{HTilde} reads
\begin{equation}\label{HT}
  \tilde{H}= \frac{\hbar}{2} \sum_{n=1}^N \omega_n(z_{0,n})\sigma_{z,n}
             + \sum_{n=1}^N \hbar\nu_n (a_n^\dagger a_n)
             - \frac{\hbar}{2}\sum_{n<l}^N J_{nl} \sigma_{z,n} \sigma_{z,l} \; .
\end{equation}
$\tilde{H}$ describes a linear string of ions with each ion
representing an  individually accessible qubit with characteristic
resonance frequency. The last term in this Hamiltonian expresses a
pairwise coupling between qubits, analogous to the well-known
spin-spin coupling in molecules used for NMR experiments. The
collection of trapped ions can be viewed as an $N$-qubit molecule
with adjustable coupling constants (compare section \ref{spin}).

\subsection{Adding a driving field}
The additional term in the Hamiltonian governing the dynamics of 
qubit $j$ when irradiated with electromagnetic radiation at 
frequency $\omega$ close to its resonance is given by
\begin{eqnarray}\label{HM}
 H_M &=& \frac{\hbar}{2} \Omega_R (\sigma_j^+ + \sigma_j^-)
        \left[\exp[i(kz_j-\omega t+\phi')]+\exp[-i(kz_j-\omega
        t+\phi')]\right] \nonumber \\
     &=& \frac{\hbar}{2} \Omega_R (\sigma_j^+ + \sigma_j^-)
        \left[\exp\left[\sum_n iS_{nj}\eta_n(\ad_n+a_n)-i\omega t + i\phi\right]
         +\mbox{h.c.}
        \right]\; .
\end{eqnarray}
First performing the unitary transformation 
$\tilde{H}_M=U^\dagger H_M U$ where it is convenient to express 
$U$ given in \ref{U} as
\begin{equation}
  U = \exp\left[
          \frac{1}{2}\sum_{n=1}^N\sum_{l=1}^N \ve_{nl}
          (\ad_n-a_n)\sigma_{z,l}
        \right] \; ,
\end{equation}
then transforming $\tilde{H}_M$ into the interaction picture
defined by
$\tilde{H}_M^I=\exp(\frac{i}{\hbar}\tilde{H}t)\tilde{H}_M
                \exp(-\frac{i}{\hbar}\tilde{H}t)$,
and finally omitting terms with time dependent factors that
contain the sum of $\omega$ and $\omega_j$ (rotating wave
approximation) gives
\begin{eqnarray}\label{HMI}
 \tilde{H}_M^I = \frac{\hbar}{2} \Omega_R
                   \left[\right. &\exp &\left[i\left(\omega_j-\omega-\frac{1}{2}
                        \sum_n\nu_n\ve_{nj}\right)t+i\phi\right]\sigma_j^+ \nonumber \\
                   &\exp&  \left[i\left(\sum_n(\eta_n S_{nj}+i\ve_{nj})\ad_n(t)
                         +(\eta_n S_{nj}-i\ve_{nj})a_n(t)
                         \right.\right. \nonumber \\
             &+&  \left.\left. i\eta_n S_{nj}\sum_l\ve_{nl}\sigma_{z,l}^{(1-\delta_{lj})}
                 \right)\right] + \mbox{h.c.}\left.\right]
\end{eqnarray}
with
 $a_n(t)=a_n\exp(-i\nu_n t)$ and $\ad_n(t)=\ad_n\exp(i\nu_nt)$.
If the driving radiation $\omega$ pertains to the rf or mw regime,
then $\eta_n$ is close to zero and the last term in the exponent
in \ref{HMI} can be neglected ($\eta_1\approx 10^{-6}$ for 10
Yb$^+$ ions with transition frequency $\omega_0 = 2\pi\,12.6$ GHz
and $\nu_1= 2\pi\, 100$kHz). With the definitions
\begin{eqnarray}\label{}
  \eta'_{nj}\exp(i \phi_j)
           & \equiv & \eta_n S_{nj} + i \ve_{nj}\; , \nonumber \\
   \phi_j  & \equiv & \frac{\pi}{2}-\tan\frac{\eta_n S_{nj}}{\ve_{nj}}
  \approx \frac{\pi}{2} \; ,\nonumber \\
\mbox{and }  \Delta_j &\equiv& \frac{1}{2} \sum_n\nu_n\ve_{nj}
\end{eqnarray}
the Hamiltonian in \ref{HMI} can be rewritten as
\begin{eqnarray}\label{HMIF}
\tilde{H}_M^I= \frac{\hbar}{2} \Omega_R \left[ \right.
                &\exp&\left[i(\omega_j +\Delta_j - \omega)t + i\phi\right]
                \sigma_j^+ \nonumber \\
                &\exp&\left[i\sum_n\eta_{nj}'(\ad_n(t)e^{i\phi_j}+
                    a_n(t)e^{-i\phi_j})\right]
                    +\mbox{h.c.} \left.\right]
\end{eqnarray}
The exact value of $\nu_n$ depends on the internal state
configuration of the ion chain. However, after summing over all
vibrational modes $\Delta_j$ in \ref{HMIF} is nearly independent
of the ions' internal states and reflects a constant shift in the
qubit's resonance frequency. The Hamiltonian \ref{HMIF} is
formally the same as the one valid for the interaction between
trapped ions and optical radiation, except that the parameter
combination $\eta_n S_{nj}$ determining the coupling strength
between external and internal dynamics has now been replaced by
the effective Lamb-Dicke parameter $\eta_{nj}' \approx \ve_{nj}$.
Any operation that requires coupling between motion and internal
dynamics and thus usually requires optical radiation can be
carried out using radiation in the rf or mw regime. For example,
conditional quantum dynamics on a collection of qubits may be
implemented according to the schemes proposed in
\cite{Cirac95,Sorensen99,Jonathan00}.

Sideband cooling is achieved when combining excitation on the
so-called red sideband resonance of an internal ionic transition
with a suitable dissipative process, similar to sideband cooling
in the optical regime. Optical sideband cooling has proven
efficient for the preparation of trapped ions close to their
motional ground state \cite{King98,Peik99,Roos99}.

\subsection{Spin resonance with trapped ions}\label{spin} The
additional spin--spin coupling term  in \ref{HT} is considered to 
be a disturbance  when schemes for quantum logic are applied -- 
specifically designed for trapped ions -- that in one way or the 
other rely on the existence of motional sidebands accompanying 
qubit transitions. The error introduced by this term is 
negligible compared to other technological limitations, and does 
not impose a new limit on the precision of ion trap quantum logic 
operations \cite{Mintert01}.

Instead of employing usual ion trap schemes, this spin-spin
coupling term may be directly used to implement conditional
dynamics using NMR methods. To obtain an order of magnitude
estimate of the coupling constant $J$ in \ref{J} we take
 $\partial_z\omega_j=(\mu_B/\hbar) \partial_z B \;\forall\; j$.
Here, state $\1$ experiences a linear Zeeman shift and there is no
shift for $\0$. This is the case, for example, with the ground
state of \Yb when $\0$ and $\1$ are identified with
 $|S_{1/2},F=0\rangle$ and $|S_{1/2}, F=1, m_F= 1\rangle$, respectively. With
 $S_{jn}\approx N^{-1/2} \approx S_{jl}$
we obtain
\begin{equation}
  J \approx \frac{1}{4N m \hbar}\left(\mu_B \frac{\partial B}{\partial z}\right)^2
  \frac{1}{\nu_1^2} \sum_{j=1}^N \frac{1}{\lambda_j^2} \; ,
\end{equation}
where $\lambda_j^2$ denotes the $j$--th eigenvalue of the
dynamical matrix $\hat{A}$. For 10 \Yb  ions, $\nu_1=2\pi\;
100$kHz, and $\partial_z B = 10$T/m,  $ J/2\pi \approx  40$Hz. The
magnitude of $J$ is comparable to values that occur in NMR
experiments where it depends on the type of molecule and nuclei
used. For example, in \cite{Jones98} $J/2\pi=7.2$Hz with protons
is quoted;  protons and carbon nuclei coupled by
$J_{HC}/2\pi=103$Hz and $J_{CC}/2\pi=201$Hz are described in
\cite{Nielsen98}; values of $J/2\pi$ ranging from 0.9Hz to 163Hz
with the same nuclei in a different molecule are reported in
\cite{Knill00}), and protons, nitrogen, carbon, and fluorine
nuclei with $J/2\pi$ between 2.7Hz and 366Hz are described in
\cite{Marx00}. Here, $J$ can be given a desired value by variation
of $\nu_1$ that characterizes the trapping potential, and of the
field gradient $\partial_z B$. If a gradient is applied that
changes with $z$, then the coupling constants $J_{nl}$ can assume
different values for different pairs of spins.

The variation of the field gradient along the $z-$axis is also
useful to simultaneously cool all vibrational modes of the ion
string \cite{Wunderlich01}.

\section{Concluding remarks}

Hitherto it was accepted that electromagnetic radiation does not
couple internal and motional degrees of freedom of trapped atoms
when long-wavelength radiation is used, since the Lamb-Dicke
parameter is negligibly small. Here, physical conditions are
described under which this coupling does occur for
electrodynamically trapped ions and can be used for QIP and other
experiments that require coherent conditional dynamics. It has
been shown that individual qubits can be distinguished by
frequency using microwave radiation. 

To date, experiments using spin resonance on the one hand and
trapped ions on the other, undoubtedly have been most successful
in the implementation of quantum computing. This proposal combines
the respective advantages of these two types of experimental
techniques:  qubits in ion traps can be individually addressed,
they are well isolated from the environment, and their number and
mutual coupling is variable over a wide range. On the other hand,
microwave and radiofrequency technology for NMR experiments has
been developed over decades. Thus a new avenue for QIP research is
opened up that may lead to simpler and more precise experimental
procedures.

Fruitful discussions with D. Rei\ss and helpful comments on the 
manuscript by W. Neuhauser are gratefully acknowledged. This work 
was supported by the Deutsche Forschungsgemeinschaft.

%

\end{document}